# Gain and charge response of 20" MCP and dynode PMTs

H.Q. Zhang[a,b,c], Z.M. Wang [a,b,c,1], F.J. Luo[b,c], A.B. Yang[a,b,c], D.R. Wu[a,b,c], Y.C. Li[b,c], Z.H. Qin [b,c], C.G. Yang[a,b,c], Y.K. Heng[a,b,c], Y.F. Wang[a,b,c] and H.S. Chen[b,d]

[a]University of Chinese Academy of Sciences,
19A Yuquan Road, Shijingshan District, Beijing, China
[b]Institute of High Energy Physics, Chinese Academy of Sciences,
19B Yuquan Road, Shijingshan District, Beijing, China
[c]State Key Laboratory of Particle Detection and Electronics,
19B Yuquan Road, Shijingshan District, Beijing, China
[d]Key Laboratory of Particle Astrophysics,
19B Yuquan Road, Shijingshan District, Beijing, China

**Abstract**

JUNO is a 20-kton liquid scintillator detector aiming to determine the neutrino mass ordering, precisely measure the oscillation parameters, detect the astrophysical neutrinos and search for exotic physics. It is designed to reach an energy resolution of 3% at 1 MeV with the highest ever PMT coverage, using two types of 20" phototubes: MCP-PMT from NNVT and dynode-PMT from Hamamatsu. In this article, the gain and charge response of the MCP and dynode PMTs are investigated with the study of JUNO Central Detector prototype. The linearity of the MCP-PMT charge output is measured too to check the effect of a long tail on its charge spectrum.

**1 Introduction**

The Jiangmen Underground Neutrino Observatory (JUNO)[1,2] is a multi-purpose neutrino experiment designed to determine the neutrino mass ordering, precisely measure the neutrino oscillation parameters, observe the astrophysical neutrinos and geo-neutrinos, and search for exotic physics. The detector will be equipped with a 20-kton liquid scintillator (LS) under over 700-meter rock burden and it will house approximately 20,000 20" photomultipliers (PMT) to reach 75% PMT coverage to achieve an energy resolution of 3% at 1 MeV[3]. A combined configuration is adopted in the JUNO design: 5,000 20" dynode PMTs (R12860-50 HQE) from Japan Hamamatsu Photonics (Hamamatsu) and 15,000 20" Micro Channel Plate (MCP) PMTs (N6201) from North Night Vision Technology Co. (NNVT)[1,23,24,27,28] . A JUNO central detector (CD) prototype was built with several types of PMTs, including the aforementioned ones[4]. Concerning the characteristics of the tested PMTs and to further understand the detector response, the calculation of PMT gain will be discussed here for the MCP and dynode PMTs.

PMT is widely used in high energy physics experiments and many other fields, where its performance on output charge is crucial and its gain should be well studied for the charge calculation. In Fig. 1, charge spectra of single photoelectron response (SPR) are shown as example for both of the 20" dynode and MCP PMTs with Earth's magnetic field (EMF) shielding, under the same configuration with an external LED pulsing in single photon level checked by Poisson distribution method [5-13], where an obvious difference between the two PMTs can be seen: the MCP-PMT has a longer tail than the dynode-PMT does which is not only from multi-photoelectron. As is well known, the number of emitted photoelectron (p.e.) from PMT photocathode follows a Poisson distribution statistically with a few input photons [5-13], which can be used to understand this difference and further determine the tube features and will be discussed in detail later. The long tail of MCP-PMT in charge spectrum, which could affect its gain and charge calculation, needs to be understood as a part of the MCP-PMT charge response for the coming usages in JUNO or other experiments [14-19]. In this article, the PMT gain's determination is reviewed for both of the dynode and MCP PMTs in section 2, and their measured charge responses are compared with toy Monte-Carlo (toy-MC) with several input light intensities in section 3. In section 4, the suggested algorithm for gain determination is given, and the corresponding PMT calibration algorithms are discussed.

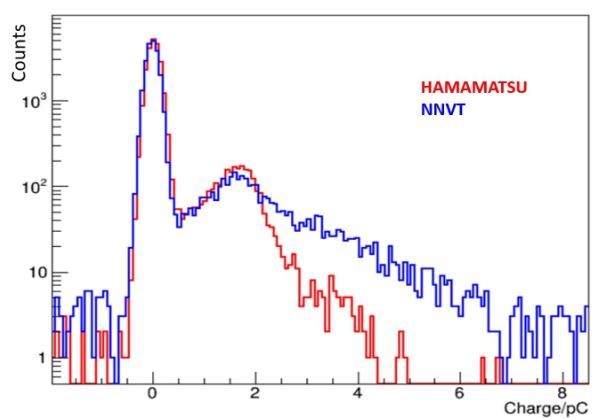

Fig. 1. Single photoelectron response (SPR) of the 20'' dynode and MCP PMTs.

**2 PMT Gain**
**2.1 Overview**
Gain is an important parameter of PMT. In the charge mode, the PMT gain, as a constant conversion factor, will directly affect the measured signal intensity in p.e. which is obtained from the output charge in pC. In the photon counting mode, the result is also affected by the PMT gain for its impact on signal to noise ratio and threshold setting. Furthermore, we need to have an equivalent gain determination algorithm for all the PMTs when different kinds of tubes are used together to avoid bias among their measured charges and to reach an equal result with photon counting method.

The PMT gain can be obtained through many ways in current or in charge. According to the manual of Hamamatsu Photonics, a current-based method of [anode luminous sensitivity] / [photocathode luminous sensitivity] is adopted[20], while another method based on SPR charge spectrum measured with pulse light source is widely used in high energy physics experiments[6,21-22] as shown in Fig.2, where the gain is determined by the single photoelectron (SPE) peak location with an external triggered and pulsed light source and data acquisition. Here, the charge of a PMT output pulse is integrated from its recorded waveform in a time window [-25,50] ns relative to its peak location 0ns where its baseline is calculated in window [-175,-100] ns relative to each pulse peak location too, and it is the same for both types of PMTs.

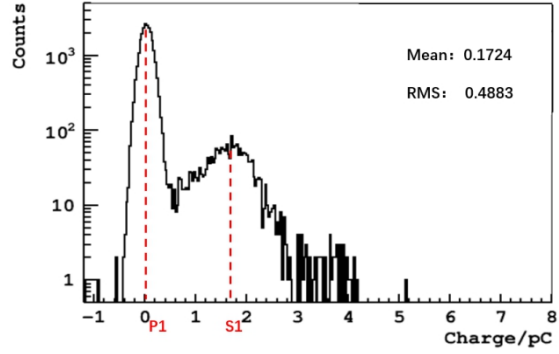

Fig. 2. The SPR with pulse light source in pC and peaks used to determine the PMT gain ($P_1$: the peak location of pedestal; $S_1$: the peak location of single photoelectron (SPE); Mean (M): averaged charge of the whole spectrum.)

Following Fig. 2, two algorithms can be used to determine the charge-based PMT gain (the calculated gain), named as $G_p$ (from the peak location ($S_1$)) and $G_m$ (from the "Mean" (M) of the whole spectrum), which are expressed in Formula (1) and Formula (2) respectively, which will only be used for gain calculation with SPR:

$$G_p = \frac{S_1 - P_1}{e} \qquad (1)$$

$$G_m = \frac{M - P_1}{e} \times \frac{1}{\mu} \qquad (2)$$

where $P_1$ and $S_1$ (in pC) are the peak locations of the pedestal and signal respectively as shown in Fig. 2, which can be fitted generally by a Gaussian function to reach less than 1% uncertainty with enough statistics. The M (Mean, in pC) is the averaged charge of the whole spectrum including the pedestal, the signal and all the other additional noises, which also can reach less than 1% uncertainty with enough statistics. The $e$ is the charge of an electron and $\mu$ is the expected mean photoelectron number following the Poisson distribution, which can reach 1% uncertainty too and will be discussed in detail later. It should be noted that formula (2) is only valid for SPR triggered and measured with pulse light source. There also are some other algorithms with similar principle as $G_p$ [6].

According to the above formulas, it is easy to see that $G_m$ is sensitive to additional noise and could encodes any abnormal response structure of the tube, while it is also related to the calculated $\mu$. On the other hand, $G_p$ is directly related to the single photoelectron peak, but ignoring possible information from abnormal structure such as the long charge tail of the MCP-PMT. We will elaborate all these factors for further understanding of the calculated gains and their differences.

**2.2 Expected mean photoelectron number**
With a constant, weak and pulsed light input and external triggered data acquisition synchronized with the light pulse, we can get the measured signal intensity by PMT in p.e. through a statistical analysis with its output charge spectrum according to Formula (3), namely the expected mean photoelectron number (denoted as $\mu$) [5,7-13]. In the formula, $N_0$ is the trigger number under the pedestal peak which is counted all the events with charge smaller than the threshold (0.25

p.e., decided by the signal noise ratio which effect will be discussed later) and $N_{trigger}$ is the total trigger number of the whole charge spectrum, which is directly proportional to the photon counting result. This method and its uncertainty are discussed in detail in [7-13], where all the listed settings need to be satisfied and $N_0$ needs to be larger than 0, which limit the usage of this method for PMT signal intensity calculation both in test configurations and intensity ranges.

$$\mu = -\ln\frac{N_0}{N_{trigger}} \tag{3}$$

With the two calculated PMT gains of Formula (1) and (2), generally we can obtain another two values of measured photoelectrons for a pulsed light signal from the PMT measured charge S (for a single event, or it is the mean of the spectrum) respectively named as $N_{pe}^{G_p}$ (Formula 4) and $N_{pe}^{G_m}$ (Formula 5), which is no other limitations (comparing with formula (3)) on test configurations or light intensities, and they can be used for all the tests and next discussions:

$$N_{pe}^{G_p} = \frac{S-P_1}{e} \times \frac{1}{G_p} \tag{4}$$

$$N_{pe}^{G_m} = \frac{S-P_1}{e} \times \frac{1}{G_m} \tag{5}$$

Where $P_1$ is the pedestal location as defined previously and S is the measured charge from each event or spectrum averaged.

Here, a 20" MCP-PMT and a 20" dynode-PMT are tested separately in detail with light input in different intensities to ensure the PMT output intensities ranging from single photoelectron to several photoelectrons. The $G_p$ and $G_m$ are calculated from the SPR spectrum (as the calibration step), which are listed in Table 1. A parameter named as $G_{ratio}$ ($G_{ratio} = G_m / G_p$) is defined to evaluate the difference between the two gains, where it is 1.60±0.03 for the MCP-PMT, which is resulted from its long tail in the charge spectrum, and it is 0.92±0.02 for the dynode-PMT. Both of the ratios are not equal to 1, the reason of which we are going to discuss further in section 3.1.

The relationship of the mentioned three algorithms by formula (3) (4) (5) for PMT output calculation, including $\mu$ from Poisson statistics, $N_{pe}^{G_p}$ and $N_{pe}^{G_m}$, is checked with the same data set. Fig. 3 presents the correlation between $N_{pe}^{G_p}$ or $N_{pe}^{G_m}$ and $\mu$ for both dynode and MCP PMTs, where a linear fitting is performed too. The fitted slope between $N_{pe}^{G_m}$ and $\mu$ is 1.02±0.00 for dynode-PMT and 1.00±0.02 for MCP-PMT, both of which are more consistent with 1 than those of between $N_{pe}^{G_p}$ and $\mu$, which is 0.94±0.00 for Hamamatsu and 1.62±0.03 for NNVT. The ratios between $N_{pe}^{G_m}$ and $N_{pe}^{G_p}$ are basically consistent with the calculated $G_{ratio}$ from SPR in Table 1. All the results show that the calculated PMT output charge in p.e. from $G_m$ is more consistent with the Poisson statistics result $\mu$ than that from $G_p$, and it is consistent between dynode and MCP PMTs. $G_m$ can be used for high light intensity calculation to get the consistent result as photon counting method as Formula 3, and it is more suitable as the suggested gain to match all the results.

Table 1. The calculated gain of EA0073 and PA1702-1560 with different methods where G$_{ratio}$= G$_m$ / G$_p$

| PMT ID | G$_p$ | G$_m$ | G$_{ratio}$ |
|---|---|---|---|
| EA0073(Hamamatsu) | (1.11±0.01)*10$^7$ | (1.03±0.01)*10$^7$ | 0.92±0.02 |
| PA1702-1560(NNVT) | (0.99±0.01)*10$^7$ | (1.61±0.03)*10$^7$ | 1.60±0.03 |

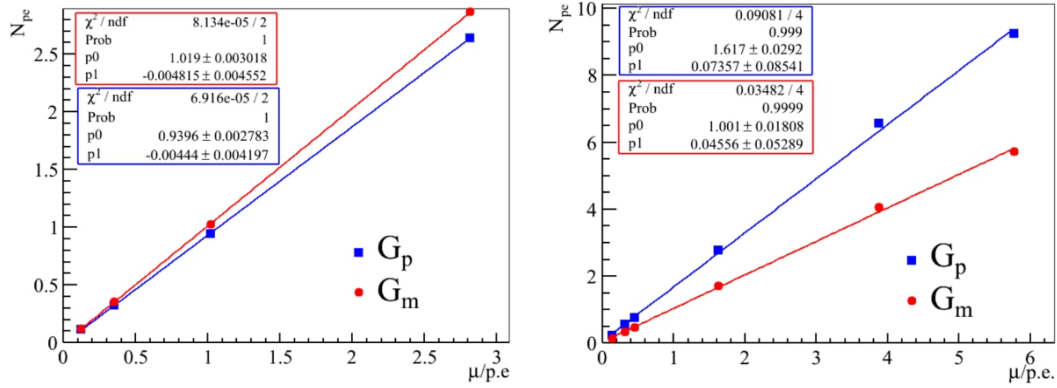

Fig. 3. The calculated $N_{pe}^{G_p}$ from G$_p$ (fitted by blue line) and $N_{pe}^{G_m}$ (fitted by red line) following Formula (4) and (5) respectively on the Y axis, where the X axis is the value calculated from the Poisson statistic method according to Formula (3). The left and right plots are the results of the dynode-PMT and MCP-PMT respectively.

**3 Toy-MC on charge response**

**3.1 Charge response of dynode-PMT**

The charge response of dynode PMT with a pulsed light source as discussed has been studied and described with a complicated function in detail [6], where totally 7 parameters are introduced as the borrowed Formula (6) trying to describe all the noise and amplification process[6,26], where $Q_0$ and $\sigma_0$ are used to describe the pedestal, $w$ and $\alpha$ are used for an exponential part, and another three parameters ($Q_1$, $\sigma_1$ and $\mu$) are used to describe the pure signal related parts. The parameters of $w$, $\alpha$ and $\mu$ would be further discussed here: $w$ is the probability of the exponential noise contribution, $\alpha$ is the coefficient of the exponential distribution, and $\mu$ is the expected mean number of photoelectrons which is proportional to the input light intensity as discussed in section 2.1.

$$S_{real}(x) = \sum_{n=0}^{\infty} \frac{\mu^n e^{-\mu}}{n!} \times \left\{ (1-w) G_n(x - Q_0) + w \times \frac{\alpha}{2} \exp\left(-\alpha(x - Q_n - \alpha\sigma_n^2)\right) \times \right.$$
$$\left. \left[ \mathrm{erf}\left(\frac{|Q_0 - Q_n - \sigma_n^2 \alpha|}{\sigma_n \sqrt{2}}\right) + \mathrm{sign}(x - Q_n - \sigma_n^2 \alpha) \times \mathrm{erf}\left(\frac{|x - Q_n - \sigma_n^2 \alpha|}{\sigma_n \sqrt{2}}\right) \right] \right\} \quad (6)$$

where

$$G_n(x) = \frac{1}{\sigma_1\sqrt{2\pi n}} \exp\left(-\frac{(x-nQ_1)^2}{2n\sigma_1^2}\right) \quad (7)$$

$$Q_n = Q_0 + nQ_1 \quad (8)$$

$$\sigma_n = \sqrt{\sigma_0^2 + n\sigma_1^2} = \begin{cases} \sigma_0 & n=0 \\ \sqrt{n}\sigma_1 & n>0 \end{cases} \quad (9)$$

With the PMT response model, we will try to scan the parameter space of $w$ and $\alpha$ with fixed PMT gain $10^7$ ($Q_1$, $G_p$) to investigate the calculated $G_{ratio}$ (Other parameters are checked too, but no big effect is observed). Here we will generate several charge spectra with $\mu$ = 0.1 p.e. (Consistent results are obtained with other value of $\mu$). An example of the generated charge spectrum is shown in Fig.4 with ($Q_0 = 0$ pC, $Q_1 = 1.602$ pC, $\sigma_0 = 0.12$ pC, $\sigma_1 = 0.44$ pC, $w = 0$, $\mu = 0.10$ p.e., $\alpha = 0.64$ pC).

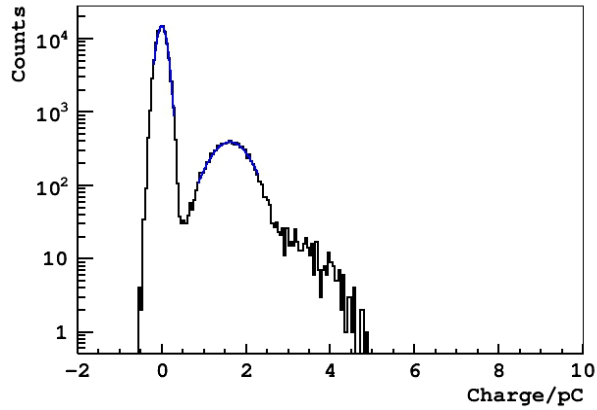

Fig. 4. An example of the generated charge spectrum following the charge response model. The pedestal and signal peak are fitted by a Gaussian function individually (blue line).

Applying the given gain algorithms in Section 2 with the charge spectra generated by toy-MC with different $w$ and $\alpha$, the calculated $G_{ratio}$s are plotted on Fig. 5, where it is clear that the value of $G_{ratio}$ can be larger or smaller than 1 and it is a function both of $w$ and $\alpha$. When $\alpha$ is smaller than 1 (specified for $G_p = 10^7$ and $Q_1 = 1.602 pC$), $G_{ratio}$ will increase dramatically following $w$'s increasing: the smaller $\alpha$, the faster $G_{ratio}$ increasing. On the other hand, when $\alpha$ is larger than 1, $G_{ratio}$ is decreasing rapidly following $w$'s increasing: the bigger $\alpha$, the faster $G_{ratio}$ dropping.

Considering the calculated $G_{ratio}$ from the 20" dynode-PMT measurement data in Section 2.2, which is 0.92±0.02, the allowed parameter space of $w$ and $\alpha$ are checked as shown on the right plot of Fig.5. Four typical configurations are shown satisfying $G_{ratio}$ = 0.92 as examples: ($w = 0.08$, $\alpha = 1.6$), ($w = 0.1$, $\alpha = 1.5$), ($w = 0.12$, $\alpha = 1.4$), ($w = 0.16$, $\alpha = 1.3$). In other words, the $G_{ratio}$ is really dependent on the PMT characteristics, and here the fixed $G_p$ is not a good candidate to fully describe PMT gain and charge response, since it ignores some features of the tube, comparing to the $G_m$, which is more suitable to describe the full response feature of the PMT and to calculate the PMT output in p.e. that is consistent with the Poisson statistics method (photon counting) as shown in Fig. 3.

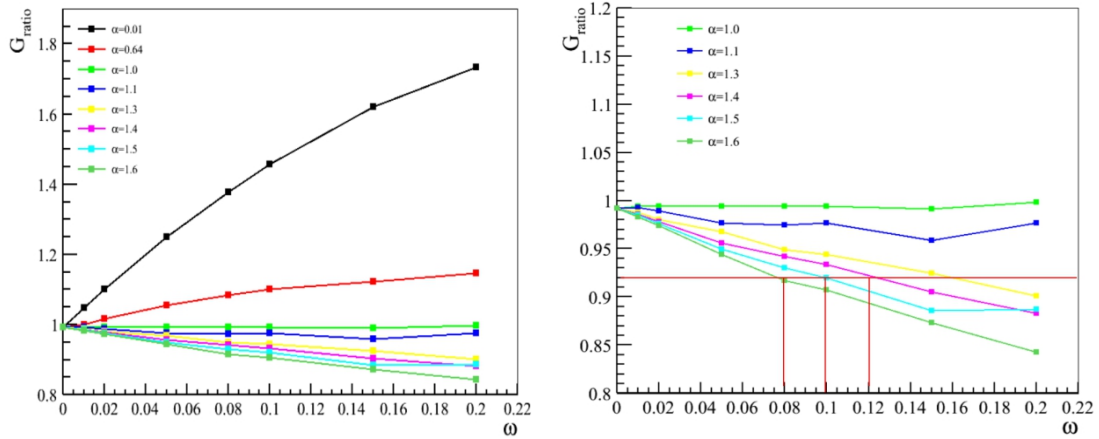

Fig. 5. Left: G$_{ratio}$ versus $w$ with different values of $\alpha$; Right: zooming in on the left plot and identifying the four configurations of $w$ and $\alpha$ which satisfies G$_{ratio}$ = 0.92 with red dash line.

**3.2 Charge response of MCP-PMT**

There is still no good charge response model for the 20" MCP-PMT as compared to the dynode-PMT, which can be used to check the gain calculation and G$_{ratio}$ as done in section 3.1. Here, we are trying to use a measured SPR of MCP-PMT output and randomly sample it to simulate the charge response of MCP-PMT. The simulated charge spectra with different input light intensities will be compared directly with the measured charge spectra after a 10 times amplifier scaled from a MCP-PMT. The PMT output intensities ($\mu$) of the measured charge spectra will be labeled by the driving voltages of the light source (LED): the higher the driving voltage, the larger the light and PMT output intensity. The pulsed photons are flashing on the full PMT photocathode with a diffuser ball. The used intensities are listed in Table 2.

The toy-MC process is divided into the following three steps:

(1) Get a measured SPR template (in pC). The measured signal intensity of the SPR has to be limited to less than 0.1 p.e. ($\mu < 0.1$). Here, a spectrum with $\mu = 0.05$ p.e. is selected. A threshold in charge at 0.21 p.e. (considering signal to noise ratio) is set to separate the spectrum into two parts as the pedestal (smaller than the threshold) and signal (larger than the threshold) which will be used in the following steps. The used SPR spectrum is displayed on Fig.6 (top-left).

(2) Generate the simulated signal spectrum in photoelectrons. Randomly sample 1,000,000 events following a Poisson distribution, where $\mu$ is set to the measured signal intensity in p.e., which is calculated from the measured charge spectrum according to Formula (3) or scaled from a reference dynode-PMT. An example of the sampled spectrum with $\mu = 0.197$ p.e. is shown in Fig. 6 (top-right).

(3) Generate the simulated charge spectrum. The PMT output is a convolution of the photoelectron emission from photocathode and the PMT charge response. Here we use the SPR template in step 1 to simulate the PMT charge response. For the photoelectron emission, the output signal n (unit in p.e., integer) will be randomly sampled from the distribution in step 2, the value of which will be used in the next step: if n = 0, randomly sample once from the pedestal part of the template in step 1 as the charge output in pC; if n >= 1, sample n times from the signal part of the template in step 1, and sum them together as the charge output in pC. The sampled charge output in pC would be filled into a histogram and further compared with the measured charge spectrum.

The comparisons of the measured and toy-MC generated charge spectra with $\mu = 0.197$ p.e. (bottom-left) and another high intensity (bottom-right) result are shown as examples in the bottom of Fig.6. Each toy-MC generated charge spectrum has 25,000 events which is the same as that in the measured charge spectrum. A parameter, $Dif_R$, is defined according to the averaged charge Q to quantify the overall difference between the measured and toy-MC generated spectra, which is expressed in formula (10) (the shape difference is discussed later). The toy-MC process is repeated 1,000 times statistically to get the averaged $Dif_R$ as in table 2. We test with all the seven PMT measured intensities covering 0-180 p.e., and the results are listed in Table 2, where all the differences are less than 1%, confirming that the measured and toy-MC generated charge spectra are consistent with each other, which also means that the MCP-PMT charge response is perfectly following the same SPR response up to at least 180 p.e. (scaled from reference dynode-PMT after over the limitation of Formula 3), and more MCP-PMT linearity studies please refer to [25].

$$Dif_R = \frac{|Q_{data} - Q_{MC}|}{Q_{data}} \times 100\% \quad (10)$$

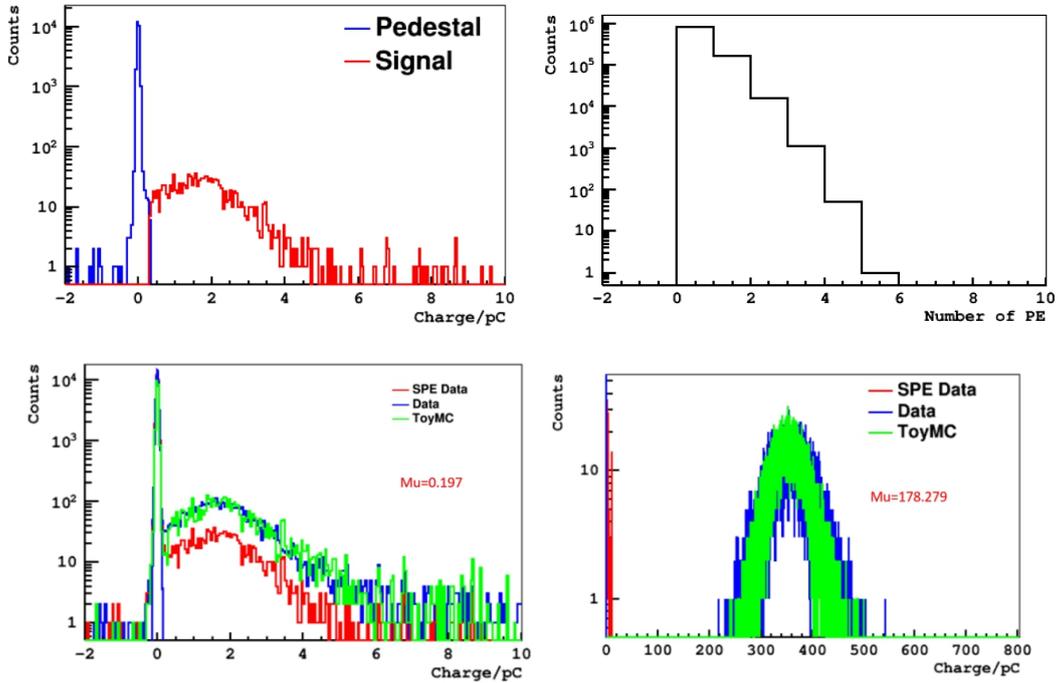

Fig. 6. Plots of the toy-MC process. Left-top: the SPE template with the pedestal and signal identified; Right-top: the distribution of N(p.e.) following a Poisson distribution with $\mu = 0.197$ p.e.; Left-bottom and right-bottom: the charge spectra comparison between measured and toy-MC results with $\mu = 0.197$ p.e. and 178.279 p.e. respectively.

While the averaged charges of the measured ($Q_{data}$) and sampled ($Q_{MC}$) spectra only give an overall information of their comparison, a further sampling test with assumed Gaussian template replacing the signal part of the template in step 1 is performed as another SPR shape related comparison. The sampled charge spectra with light intensity at 0.197 p.e. are shown in Fig.7. The difference between the sampled and measured spectra is defined by:

$$\frac{\chi^2}{ndf} = \frac{\sum_i \frac{(X_{data}^i - X_{MC}^i)^2}{X_{data}^i}}{ndf} \tag{11}$$

where $X_{data}^i$ and $X_{MC}^i$ correspond to the event count of each bin in the measured and sampled charge spectra respectively, and ndf is the number of bins. The calculated results are listed in Table 2 too. It can be concluded that the sampled spectrum based on SPR template is more consistent with measured data than that from Gaussian template especially with low light intensities.

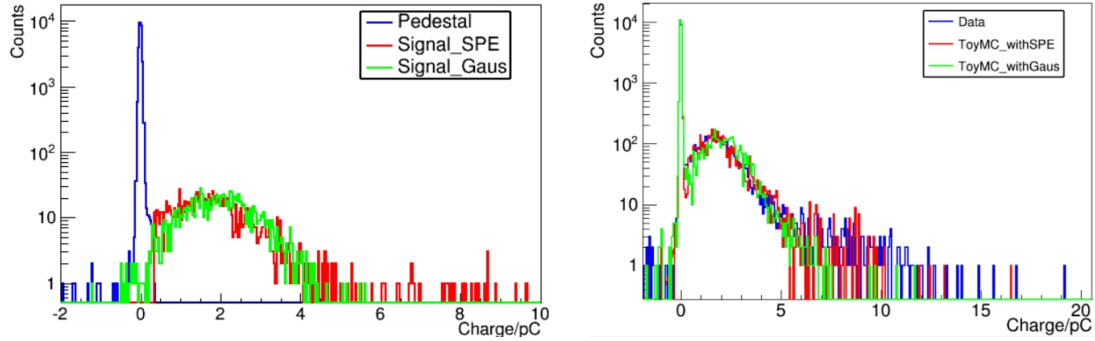

Fig. 7. Left: the template of the pedestal and signal (signal$_{SPE}$ from data and signal$_{Gaus}$ from hypothesis); Right: the sampled charge spectra with light intensity at 0.197 p.e. with signal$_{SPE}$ and signal$_{Gaus}$ compared to the measured spectrum.

Table 2. Dif$_R$ and $X^2$/ndf (from SPE and Gaussian sampling) versus PMT signal intensity (in p.e.) for the MCP PMT

| LED Driving Voltage /V | Output intensity /p.e. | Dif$_R$ (%) | SPR $X^2$/ndf | Gaussian $X^2$/ndf |
|---|---|---|---|---|
| 2.85 | 0.197 | 0.1 | 1.60 | 3.41 |
| 2.90 | 0.717 | 0.4 | 1.75 | 4.63 |
| 2.95 | 2.402 | 0.8 | 1.73 | 3.41 |
| 3.00 | 7.652 | 0.8 | 2.23 | 2.81 |
| 3.05 | 23.899 | 0.1 | 1.90 | 2.16 |
| 3.10 | 70.385 | 0.6 | 1.38 | 1.36 |
| 3.15 | 178.280 | 0.0 | 0.91 | 1.08 |

**3.3 Suggested PMT gain algorithm**

In order to understand the PMT gain and response better, we further investigate the MCP-PMT charge response here by reusing the example spectra with high light intensity in section3.2, as shown in Fig.8. The $G_p$ and $G_m$ are extracted from the SPR template to be $1.00*10^7$ and $1.25*10^7$ respectively, and the averaged signal intensity used in the Poisson distribution sampling of toy-MC is 178.3 ±0.1 p.e., referring to a dynode reference PMT in linearity measurement.

With the toy-MC, the simulated charge spectrum shows that the mean charge is 354.7±0.2 pC, which is calculated to be 177.1±0.6 p.e. with $G_m$, while if with $G_p$, it is 221.4±0.6 p.e., which is far away from the input signal intensity. In summary, the charge output in p.e. of the MCP-PMT needs to be calculated by $G_m$. These numbers are summarized in Table 3, where it can be seen that $G_m$ provides a more consistent result with the toy-MC input signal intensity than $G_p$ does.

Besides, we also have done the comparison with toy-MC between $G_p$ and $G_m$ for the dynode-PMT, the result is similar as shown in Fig.9.

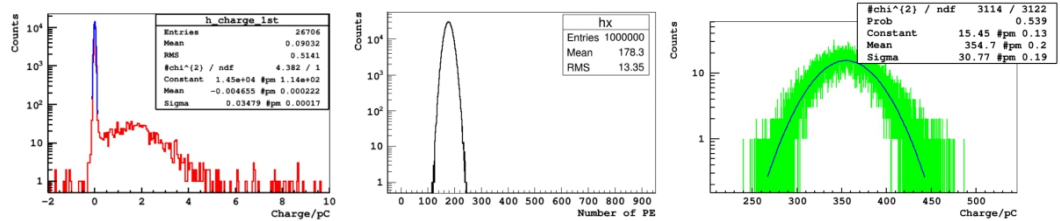

Fig. 8. An example of MCP-PMT charge spectrum with toy-MC input signal intensity of 178.3±0.1 p.e.. Left: the SPR template; Middle: the sampling with Poisson distribution with mean = 178.3±0.1 p.e.; Right: the simulated charge spectrum with toy-MC compared to the the measured spectrum.

Table 3. The expected intensity (in p.e.) calculated with $G_p$ and $G_m$ and compared with the toy-MC input signal intensity (in p.e.) for the MCP-PMT

| Method | Gain | Input Mean/p.e | Expected charge/p.e |
|--------|------|----------------|---------------------|
| $G_p$  | 1.00E7 | 178.3+/-0.1 | 221.4+/-0.6 |
| $G_m$  | 1.25E7 | 178.3+/-0.1 | 177.1+/-0.6 |

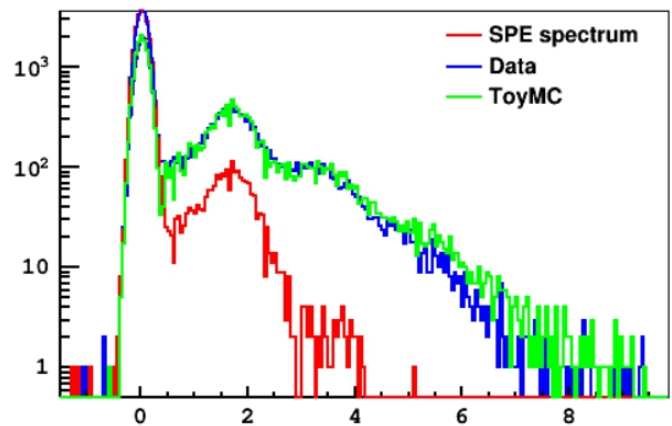

Fig. 9. The charge response of the dynode-PMT

**4 PMT gain calibration**

PMT gain is better to be determined by $G_m$ as discussed, which also requires a suitable calibration algorithm. The algorithms with light source (LED or Laser) or dark noise are commonly used for gain calibration, where the light source method is usually used, named as external trigger and can get the gain with Formula (1) (2), while the dark noise method could be configured in two different ways, named as discriminator and random trigger, which normally only can be used with Formula (2).

All the ways are compared with our measurements for both of the dynode and MCP PMTs in a sea-level laboratory, and the measured SPR charge spectra from three configurations are depicted in Fig.10, where the charge spectra from the dark noise are normalized to that of the light source method according to the height of SPE peak.

Based on the measured results shown in Table 4 and Table 5, the single electron peak's location of the dark noise or light source method are consistent within error for both of the MCP and dynode PMTs, while the averaged charges with or without threshold between the light source and dark noise methods all show big differences. The spectra measured with dark noise by discriminator or random trigger methods all have a long tail comparing with the spectrum with light source method for both of the 20" dynode and MCP PMTs, which will generate bias to the PMT gain calculation for $G_m$ calculation, and similar effect from the threshold cutoff of the discriminator method, while the spectrum with the light source method is more suitable for $G_m$ calculation as defined. We suggest that the calibration with light source and external trigger would be the most reasonable method for $G_m$ determination. While, $G_p$ from the dark noise method is a good solution to monitor the PMT gain.

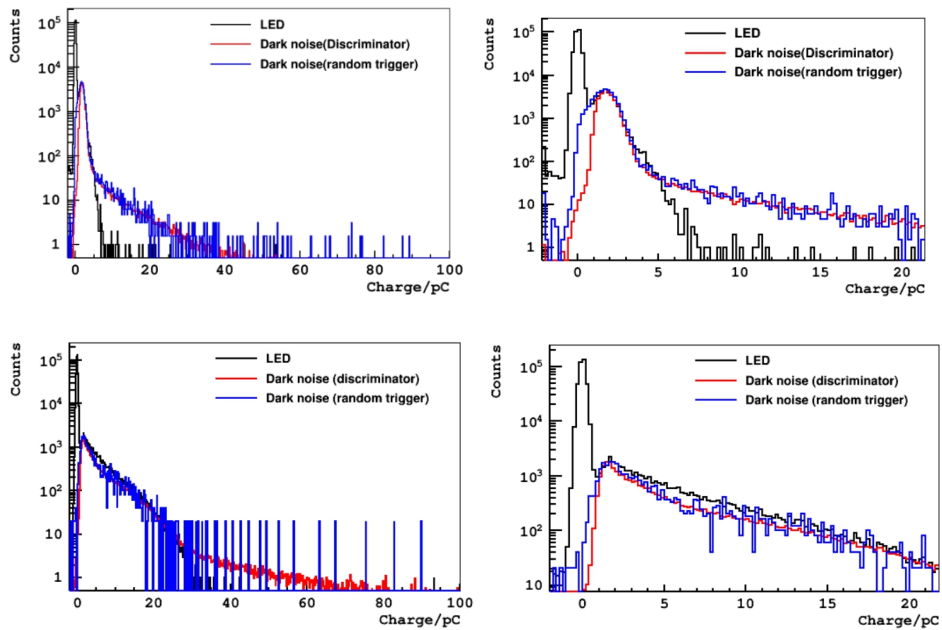

Fig. 10. The SPR charge spectra from different measurement configurations for the dynode (top) and MCP (bottom) PMTs, where the right pictures are zoomed in of the left pictures.

Table 4. The peak locations (in pC) of the MCP and dynode PMTs' charge spectra with three configurations of gain calibration: LED with external trigger, discriminator with 2mV threshold in data taking and random trigger with 2mV threshold in offline analysis.

| Method | MCP | Dynode |
| --- | --- | --- |
| LED w/ external trigger | 1.71 $\pm$ 0.01 | 1.68 $\pm$ 0.01 |
| Discriminator w/ 2mV threshold | 1.68 $\pm$ 0.01 | 1.71 $\pm$ 0.01 |
| Random trigger w/ 2mV threshold | 1.77 $\pm$ 0.04 | 1.72 $\pm$ 0.03 |

Table 5. The averaged charges (in pC) of the MCP and dynode PMTs' charge spectra with three calbration configurations: LED (with or without 0.25 p.e. threshold in offline analysis),

discriminator (with 2mV threshold in discrimnator or 0.25 p.e. threshold in offline analysis) and random trigger (with 2mV or 0.25 p.e. threshold in offline analysis).

| Method | MCP-PMT | Dynode |
| --- | --- | --- |
| LED w/ external trigger and w/o threshold | 0.56 ± 0.00 | 0.19 ± 0.00 |
| LED w/ 0.25 p.e. threshold | 4.48 ± 0.02 | 1.65 ± 0.01 |
| Discriminator w/ 2mV threshold | 6.23 ± 0.01 | 2.43 ± 0.01 |
| Discriminator w/ 0.25 p.e. threshold | 6.24 ± 0.01 | 2.44 ± 0.01 |
| Random trigger w/ 2mv threshold | 5.87 ± 0.23 | 2.02 ± 0.02 |
| Random trigger w/ 0.25 p.e. threshold | 5.94 ± 0.23 | 2.11 ± 0.02 |

**5 Discussion**

As discussed, $G_p$ can be a robust variable for PMT monitoring and calibration but its calculation shows bias to the photon counting method result based on Poisson statistics, while $G_m$ is consistent with photon counting and can be used for higher light intensity but is easily affected by dark noise. The effect of dark noise can be calculated for $\mu$ and the mean of the charge spectrum for uncertainty estimation. For example, with a 30 kHz dark noise rate and 100 ns time window for PMT pulse charge integration, the dark noise coincidence ratio to signal is ~0.003 ($\Delta\mu$). The corresponding contribution to the mean output charge is ~0.003 p.e. (ΔMean), assuming a dark noise charge of ~1 p.e.. The resulting effect is < 3% in maximum even with the minimum calibration light intensity ($\mu$) of 0.1 p.e., and the higher light intensity, the smaller effect.

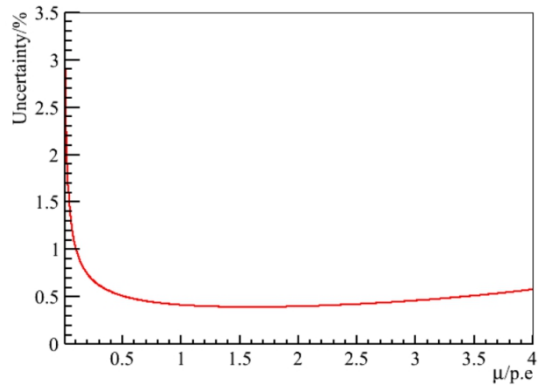

Fig. 11. The uncertainty of $\mu$ calculation vs. the light intensity following formula (3)

A calculation is done for the light intensity effect based on Formula (3) as shown in Fig. 11 and in[7]. It will be better than 1% uncertainty if the signal intensity is higher than 0.1 p.e. with enough statistics.

Concerning the threshold effect in the $\mu$ calculation in Formula (3), it is calculated with different thresholds ($0.25 \pm 0.02$ p.e.) at the noise level < 0.1 p.e. (sigma of the pedestal), the single photoelectron charge resolution of ~35% (sigma of the Gaussian) and the signal intensity of ~0.5 p.e., the total effect is less than 1%.

Finally, considering all the related effects, $G_m$ can be calculated with better than 2% uncertainty with a signal intensity of ~0.5 p.e., a 1/4 p.e. threshold and enough statistics.

## 6 Conclusion

In this paper, we investigated the PMT gain determination for both dynode and MCP PMTs, and compared their charge responses between the measured and toy-MC simulated charge spectra. According to the analysis, $G_m$ is extracted from the entire charge spectrum and compared with $G_p$, which is extracted from the peak location of the single photoelectron, for both types of PMTs. The $G_{ratio}$ ($G_m/G_p$) of the 20" Hamamatsu dynode-PMT is smaller than 1 because of the incomplete amplification and related noise, while the $G_{ratio}$ is larger than 1 of the MCP-PMT because of its long charge tail. It is also concluded that the PMT charge output in p.e. calculated by $G_m$ is more consistent with the result from photon counting method than that from $G_p$ for both types of PMTs, which indicates that $G_m$ should be considered for the future charge and energy reconstruction if it is based on PMT charge. It is also confirmed that the MCP-PMT charge output is linear to the input light intensity as the traditional dynode-PMT even with a long tail in its charge spectrum. Moreover, the light source with external trigger data acquisition is suggested for $G_m$ calibration and the $G_p$ of single photoelectron peak location from dark noise with hardware discriminator or random trigger may be used for monitoring the PMT gain.

**Acknowledgements**

This work has been supported by the National Natural Science Foundation of China (Grant No. 11875282), and the Strategic Priority Research Program of the Chinese Academy of Sciences (Grant No. XDA10010200, No.XDA10010300, No.10011100). We also thank Haoqi Lu and Zeyuan Yu for the valuable discussion on toy MC.